\documentstyle[aps]{revtex}
\def\be{\begin{equation}}
\def\ee{\end{equation}}
\def\bea{\begin{eqnarray}}
\def\eea{\end{eqnarray}}
\begin{document}  

\title{Grand Unification of the Sterile Neutrino}

\author{Biswajoy Brahmachari and Rabindra N. Mohapatra}
\address{\it Department of Physics, University of Maryland,
College Park, MD 20742, USA }

\maketitle

\begin{abstract}

The simplest way to simultaneously understand all existing indications of 
neutrino oscillations from solar and atmospheric neutrino deficits and 
the LSND experiment, seems to be to postulate a sterile neutrino. We 
present a realistic grand unified model based on the gauge group 
$SO(10)\times SO(10)^\prime$ that leads to the desired masses and mixings
for the sterile and the known neutrinos needed to understand the above 
observations while fitting those of the known charged fermions. The model 
is a grand unified realization of the recently 
proposed idea that the sterile neutrino is the lightest neutrino of a 
mirror sector of the universe which has identical matter and gauge 
content as the standard model. The two $SO(10)$'s operate on the two 
sectors in a mirror symmetric way and  are connected by a mixed Higgs 
representations whose net effect is to connect the superheavy right 
handed neutrinos of the two sectors.
\end{abstract}

\hskip 6cm UMD-PP-98-118
 
\section{Introduction}\hspace{0.5cm} 

A turning point in the search for new physics beyond the standard model
may be at hand in the arena of neutrino physics. The standard model predicts
zero neutrino masses but there are now strong indications for
neutrino oscillations (and hence nonzero neutrino masses) from the five 
solar neutrino experiments (Kamiokande, Homestake, Gallex, Sage and
Super-Kamiokande\cite{expt,superK}), four atmospheric neutrino observations
\cite{superK,atmos,fukuda} and the direct laboratory observation 
in the LSND experiment\cite{LSND}. Furthermore, to explain all three 
experiments, three 
different scales for mass differences ($\Delta m^2$) are needed.
 Since with the three known neutrinos one can get only 
two independent $\Delta m^2$'s, it has been suggested\cite{caldwell} that 
a fourth sterile neutrino be invoked. In the presence of this extra 
neutrino species ( $\nu_{s}$), one can construct several 
scenarios for solving the three neutrino 
puzzles\cite{caldwell,giunti,foot,smirnov}. In this letter, we will be 
interested in
the scheme\cite{caldwell}, where the solar neutrino puzzle is solved via
the oscillation of the $\nu_e$ to $\nu_s$ using the MSW 
mechanism\cite{msw} and the atmospheric neutrino puzzle is solved 
via the $\nu_{\mu}-\nu_{\tau}$ oscillation. The solar neutrino puzzle 
fixes the $\Delta m^2_{e-s}\simeq (0.3-1.0)\times 10^{-5}$ eV$^2$, 
whereas the atmospheric neutrino puzzle implies that $\Delta 
m^2_{\mu-\tau}\simeq 10^{-2}-10^{-3.5}$ eV$^2$ and the LSND experiment 
keeps the $0.3~eV^2 \leq \Delta m^2_{\nu_e-\nu{\mu}}\leq 10$ eV$^2$. In 
this picture the $\nu_{\mu}$ and $\nu_{\tau}$ are nearly degenerate and
near maximally mixed. If the existence of the sterile neutrino becomes 
confirmed say,
indirectly by KARMEN\cite{drexlin} observing $\nu_{\mu}-\nu_e$
oscillation or directly by SNO neutral current data to come in the
early part of the next century, a key theoretical challenge will be to
construct an underlying theory that embeds the 
sterile neutrino along with the others with appropriate mixing pattern,    
while naturally explaining its ultralightness.
 In this letter, we propose a grand unified model
which not only explains the smallness of neutrino masses via the seesaw 
mechanism but it also incorporates the sterile neutrino, whose ultralight 
mass is naturally explained. The model at the same time predicts mass 
degeneracy between maximally mixed $\nu_{\mu}$ and $\nu_{\tau}$.

The underlying framework for the present work will be the suggestion that 
there is a parallel standard model\cite{bere,foot,blini} which is an exact 
copy of
the known standard model (i.e. all matter and all gauge forces identical).
The two ``universes'' communicate only via gravity or other forces that are
equally weak. At an overall level, such a picture emerges quite naturally in 
superstring theories which lead to $E_8\times E_8^\prime$ gauge theories 
below the Planck scale with both $E_8^\prime$ s connected by gravity. We 
hasten to 
emphasize that despite this apparent promising connection, no vacuum 
state that leads to the details needed for our neutrino model has 
been discussed to date. In this paper, we will assume the sub-Planck GUT 
group to be a subgroup of $E_8\times E^\prime_8$ in the hope of possible 
future string embedding of our model. For alternative theoretical
models for the sterile neutrino, see Ref.\cite{smir}.

As suggested in Ref.\cite{bere}, we will assume that the process of 
spontaneous symmetry breaking introduces asymmetry between the two universes 
e.g. the weak scale $v^\prime_{wk}$ in the 
mirror universe is larger than the weak scale $v_{wk}= 246$ GeV in our 
universe. It was shown in Ref.\cite{bere} 
that with this one assumption alone, the 
gravitationally generated neutrino masses\cite{ellis} can provide a 
resolution of the solar neutrino puzzle
(i.e. one parameter generates both the required $\Delta m^2_{e-s}$ and the
mixing angle $sin^22\theta_{e-s}\simeq 10^{-2}$). 

In this paper, we construct a complete realistic model for known 
particles and forces and
make detailed numerical predictions for the neutrino sector within a 
grand unified scheme that implements the seesaw mechanism. Since the 
simplest GUT model that implements the seesaw mechanism is based on the 
$SO(10)$, we use SUSY $SO(10)\times SO(10)^\prime$ as our gauge group with 
each $SO(10)$ operating in one sector. As is well known, the SO(10) model
generally predicts a hierarchical pattern for all fermion masses including
neutrinos. However the LSND results in conjunction with the atmospheric 
neutrino data imply that $m_{\nu_{\mu}}\simeq m_{\nu_{\tau}}$. The 
challenge is therefore to construct an SO(10) model which will lead to
the above non-generic prediction.

We impose a mirror symmetry\cite{volkas} between the two SO(10) 
sectors so that the field contents as well as the Yukawa couplings in the 
two sectors are identical to each other and all differences between 
them arise from the process of spontaneous symmetry breaking. In 
order to constrain the model further, we impose 
the requirement that it conserve R-parity automatically without using 
any extra global symmetries. This ensures that there is a natural cold 
dark matter candidate in the model. We also impose an additional global
permutation symmetry $S_3$ which plays a key role in ensuring the mass 
degeneracy between the tau and muon neutrinos. This leads to a 
three neutrino texture which is similar in form to the one discussed in 
Ref.\cite{satya}.
The connection between the visible and the mirror sector occurs via the 
mixing of the heavy right-handed neutrinos\cite{collie}. 

 The nontrivial nature of the problem arises from the 
fact that in a GUT framework the neutrino couplings are intimately linked 
to the charged fermion couplings and it is by no means obvious that with 
a simple set of Higgs fields one can make the observed hierarchical 
pattern of the charged fermion masses and mixings compatible with the 
apparent non-hierarchical mass and mixing pattern for the neutrinos.

\section{ The model: each sector} \hspace{0.5cm}

The fermions of each generation are assigned to the ${\bf (16, 1)\oplus (1, 
16^\prime)}$ representation of the gauge group. We denote them by 
$\Psi_{e,\mu,\tau}$ in the visible sector and by corresponding symbols 
with a prime in the mirror sector (as we do for all fields). The $SO(10)$
symmetry is broken down to the left-right symmetric model by the combination
of ${\bf 45\oplus 54}$ representations in each sector. The $SU(2)_R\times 
U(1)_{B-L}$ gauge symmetry in turn is broken by the ${\bf 126 \oplus 
\overline {126}}$ representations and we take three such representations 
(and denote them by $\Delta_{0,1,2} \oplus \overline {\Delta}_{0,1,2}$). The 
role of the these fields is two-fold: First, they guarantee automatic 
R-parity conservation and second, they lead to the see-saw 
suppression for the neutrino masses\cite{seesaw}.

The standard model symmetry is then broken by the {\bf 10}-dim. Higgs 
fields of which we take three $H_{0,1,2}$. As is well-known, the {\bf 
126}-dim. representation contains in it left-handed triplets with $B-L= 
2$. Due to the presence of the {\bf 54}-Higgs field $S$ in the model, 
couplings of type $\overline \Delta \overline \Delta S$ and $H H S$ are 
allowed and they lead to induced $B-L$ breaking vev's $v_L$ which give a 
direct Majorana mass to the neutrinos leading to the so called type II 
see saw formula\cite{goran} written symbolically as
\be
m_{\nu}\simeq fv_L - \frac{m^2_{\nu^D}}{fv_R}
\ee 
where $v_R$ is the generic vev of the $\nu^c\nu^c$ component of {\bf 126}.
As was shown in \cite{goran}, the detailed minimization of the potential 
in such theories leads to the conclusion that $v_L$ is also suppressed by 
a see saw like formula (i.e. $v_L\simeq v^2_{wk}/\lambda v_R$ where 
$\lambda$ is an unknown parameter in the superpotential). If we choose
$v_R\simeq 10^{14}-10^{15}$ GeV (so that it is not far from the GUT scale)
and $\lambda \simeq 0.1-0.01$, then we get $v_L$ in the eV range. Note that
while the second term in Eq. (1) arising from the conventional see saw 
formula leads to a hierarchical mass pattern for the neutrinos, the 
first term has no such obligation. Thus, if we require some neutrinos to 
be nearly degenerate, the first term has to be given the dominant role
as we do here.

A second point is that in the effective MSSM derived from the model, the
low energy Higgs doublets will be assumed to be linear combinations of
the doublets present in all {\bf 10} as well as {\bf 126} dimensional
multiplets. In principle this situation can be realized by appropriate
arrangement of parameters.

Next we assume the invariance of the action under a discrete permutation 
symmetry $S_3$ under which the $(\Psi_{\mu}, \Psi_{\tau})$, 
$(\Delta_1,\Delta_2)$ and $(H_1,H_2)$ transform as doublets whereas 
$\Psi_e$, $H_0$ $\Delta_0$ and the rest of the fields transform as 
singlets. The
same discrete operates in the mirror sector (i.e. no mirror version of
$S_3$). This then restricts the form of the Yukawa part of the superpotential
to the following form:
\bea 
&& W_Y = h_1 \Psi_e \Psi_e H_0 + h_2~ 
(\Psi_{\mu}\Psi_{\mu} + \Psi_{\tau} \Psi_{\tau})~H_0 
 + h_3~\Psi_e (\Psi_{\mu}~H_1+\Psi_{\tau}~H_2)
+h_4~[(\Psi_{\mu}~\Psi_{\mu}-\Psi_{\tau}~\Psi_{\tau})~H_1
\nonumber\\ 
&& +2~\Psi_{\mu}~\Psi_{\tau}~H_2]
+ f_1~\Psi_e~\Psi_e~\overline{\Delta}_0 + 
f_2~(\Psi_{\mu}\Psi_{\mu}+\Psi_{\tau}\Psi_{\tau})~\overline{\Delta}_0  
 +f_3~[(\Psi_{\mu}\Psi_{\mu}-\Psi_{\tau}\Psi_{\tau})\overline{\Delta}_1
+2\Psi_{\mu}\Psi_{\tau} \overline{\Delta}_2] \nonumber\\
&& + f_4~\Psi_e~(\Psi_{\mu}\overline{\Delta}_1 
+\Psi_{\tau}\overline{\Delta}_2)
\eea
Using Eq. 2, we can write down the quark and lepton mass matrices for 
the visible sector as follows.
\be
M^D_{10} = \pmatrix{ c_1 & c_2 & c_3 \cr
              c_2 & c_4+c_5 & c_6 \cr
              c_3 & c_6 & c_4-c_5}
~~~;~~~
M^D_{126} = \pmatrix{d_5 & d_1 & d_2 \cr
                   d_1 & d_3+d_6 & d_4 \cr
                   d_2 & d_4 & d_6-d_3}
\ee   
\be   
M^U_{10} = \pmatrix{a_1 & a_2 & a_3 \cr
                  a_2 & a_4+a_5 & a_6 \cr
                  a_3 & a_6 & a_4-a_5}
~~~;~~~
M^U_{126} = \pmatrix{b_5 & b_1 & b_2 \cr
             b_1 & b_3+b_6 & b_4 \cr
             b_2 & b_4 & b_6-b_3}
\ee
\be
M^U = M^U_{10}+M^U_{126}~~~;~~~M^D = M^D_{10}+M^D_{126}~~~;      
~~~M^\nu = M^U_{10} - 3 M^U_{126}~~~;~~~M^L = M^D_{10} - 3 M^D_{126}
\ee
Even though apriori, it may appear from Eq. 3 and 4 that there are 24 
parameters in the visible sector mass matrices, the actual number is 16 
due to the $S_3$ invariance of the theory which yields eight relations 
among them. They are  
$\frac{c_3}{c_6}=\frac{c_2}{c_5}=\frac{a_2}{a_5}=\frac{a_3}{a_6} $; 
$\frac{a_1}{a_4} =\frac{c_1}{c_4}$; 
$\frac{b_1}{b_3}=\frac{d_1}{d_3}=\frac{b_2}{b_4}=\frac{d_2}{d_4}$ and
$\frac{b_5}{b_6}=\frac{d_5}{d_6}$.
We now proceed to determine the remaining parameters in such a way
they give rise to observed fermion masses and quark mixings and vanishing
charged lepton mixings. The best fit is obtained for the following values
for standard model parameters at the GUT scale
 (all masses in GeV units): $m_t=112.00, m_c=0.370,
m_u=0.0011$; $ m_b=1.115, m_s = 0.0148, m_d=0.0013$; CKM mixing parameters
$s_{12}=0.2201; s_{23}=0.031; s_{13}=0.0039 $ and the lepton masses
$m_e=0.00033, m_{\mu}=0.0699, m_{\tau}=1.1817$ we find the values
for the parameters $a,b,c,$ and $d$ listed in table I. They in turn enable 
us to determine the Dirac neutrino mass matrix for
the visible as well as the mirror sector. Although the Dirac mass 
matrix for the neutrino does not play a significant role in the masses
and mixings in the individual sector, we will see in the next section that 
it plays a crucial role in the mixing between the two sectors.

\section{Connecting the two sectors and predictions for neutrinos}

The neutrino mass matrix has two contributions as is seen from Eq. 1. 
For $v_R$ near $10^{15}$ GeV, the largest entry from the second term in
Eq. 1 is of order $10^{-2}$ eV in the $33$ element and much smaller in 
other places. As far as the first term goes, its form is dictated by Eq. 2
and we choose it as follows:
\be
M_{\nu\nu}=\pmatrix{0 & A_l & A^\prime_l \cr
                    A_l & B_l & D_l \cr
                    A^\prime_l & D_l & -B_l}~~{\rm in~eV}
\ee
Where ${A \over A^\prime}={B \over D}$. This mass matrix looks very 
similar to the one analyzed in \cite{satya}. 
In order to obtain the predictions for neutrino masses and mixings, we 
need to know the structure of the mass matrices in the mirror sector and 
the connection between the visible and the mirror sector.

The exact mirror symmetry between the visible and the mirror sector implies
that at the level of the superpotential, all couplings in the mirror sector
are identical to those in the visible sector. We will assume that the 
spontaneous symmetry breaking breaks the mirror symmetry so that actual 
mass matrices will exhibit differences. For simplicity, we will assume that 
all doublet vev's in the mirror sector differ by a common ratio from those
in the visible sector (i.e. $v^\prime_{wk}/v_{wk}=\zeta$ ).
Since this asymmetry will effect the fermion masses in the two MSSM's, we
will expect the $B-L$-breaking scales and the GUT scales to be different.
This in turn will imply that the induced triplet vev's will also be different
in the two sectors. Our strategy will therefore be to scale the Dirac mass
matrix for the mirror sector by a common factor but introduce arbitrary
triplet vev's in the mirror sector. 
\be
M_{\nu^\prime \nu^\prime }=\pmatrix{\alpha & A_m & A^\prime_m \cr
                       A_m & B_m & D_m \cr
                       A^\prime_m & D_m & -B_m}~~{\rm in~eV}
\ee
Where, $A_m= q_1 A_l$, $A^\prime_m= q_1 A_l$, $D_m= q_2 D_l$ and $B_m=q_2 
B_l$. Let us now try to connect the two sectors which we do by 
postulating the
Higgs fields ${\bf (16, 16^\prime)\oplus (\bar{16}, \bar{16^\prime})}$ 
(denoted by $\chi\oplus \bar{\chi}$). There can now be a connecting term 
between the two sectors given by 
\be
W^\prime = g_c\Psi_e\Psi^\prime_e\bar{\chi} 
+g^\prime_c(\Psi_{\mu}\Psi^\prime_{\mu}
+\Psi_{\tau}\Psi^\prime_{\tau})\bar{\chi}
\ee
We now give a vacuum expectation value to the $\nu^c{\nu^c}^\prime$ element
of $\chi\oplus\bar{\chi}$ fields. Then only the right handed neutrinos
of both sectors get connected. This in conjunction with the Dirac masses
of both sectors introduces a mixing matrix between the two sectors of the 
following form (assuming for simplicity $g_c=g^\prime_c$): 
\be
M_{\nu\nu^\prime} =g_c M_{\nu^D}M^{-2}_{\nu^c}M_{\nu^D}<\chi> 
\ee
Where,
\be
M_{\nu^c \nu^c}=\pmatrix{0          &  A_r  & A^\prime_r \cr
                   A_r    &  B_r  & D_r \cr
               A^\prime_r &  D_r  & -B_r} ~~~{\rm in~GeV}
\ee
Where, $ A_r= l_1 A_l$, $A^\prime_r= l_1 A^\prime_r$, $B_r = l_2 B_l$ 
and $D_r = l_2 D_l$. We diagonalize the complete neutrino mass matrix to 
obtain the following absolute values of the mass eigenvalues (in eV's):
$m_{\nu^\prime_\tau}=90.56, m_{\nu^\prime_\mu}=-90.56, 
m_{\nu^\prime_e}=0.0034$ $ m_{\nu_\tau}=1.51, m_{\nu_\mu}=-1.509, 
m_{\nu_e}=0.001 $. The squared mass diffences (in eV$^2$) are $\Delta 
m^2_{e-s}= 9.9 \times 10^{-6}$, 
$\Delta m^2_{\mu-\tau}=0.003$ and $\Delta m^2_{e-\mu}= 2.27$, where the 
numbers are given in eV$^2$. The fitted values of the parameters are, 
$\alpha=0.005$, $A_l=0.0253$, $A^\prime_l=0.050$, $B_l=0.675$, 
$D_l=1.35$ given in eV 
units, $q_1=10$, $q_2=60$, $l_1=1.2~10^{15}$, $l_2=0.5~10^{15}$ and 
$g_c<\chi>=6.9\times 10^{12}$ given in GeV units. The mixing matrix 
$O^\nu$ of the six neutrinos in the basis ($\nu_e,\nu_{\mu},\nu_{\tau}, 
\nu^\prime_e,\nu^\prime_{\mu},\nu^\prime_{\tau}$) 
is approximately given as,
{\small
\be 
O^{\nu} = \pmatrix{-0.99 & 0.037 & 0 & 0.039 & -0.00025 & 0 \cr 
   -0.031 & -0.85 & -0.52 & -0.00072 & 0 & 0 \cr 
   0.019 & 0.525 & -0.85 & -0.00043 & 0 & 0 \cr 
   -0.042 & 0.0014 & 0.00071 & 
    -0.999 & 0.0062 & 0 \cr 
   0 & 0 & 0 & -0.0053 & -0.850 & -0.525 \cr 
   0 & 0 & 0 & 0.0032 & 0.52 & -0.85}
\ee
}
Combining this with the mixing angle for the charged leptons, we obtain 
the final mixing matrix among the four neutrinos which looks identical 
to the corresponding top-left $4\times 4$ submatrix of $O_{\nu}$ with only 
the $\nu_e-\nu_{\tau}$ entry reduced by a factor of 2 because of the 
presence of a small 13 element in the charged leptonic mass matrix. 
We have varied the 
parameters of the model to see the allowed range for the $m_{\nu_{\mu}}$
relevant for the LSND experiment and find consistent solutions for the
range $0.5\leq m_{\nu_{\mu}}/eV \leq 1.5$. Therefore, the $\nu_{\mu}$ and
$\nu_{\tau}$ together could play the role of the hot dark matter for
the upper allowed range of the masses. Also note that the zeros in the
neutrino mixing matrix simply means that those entries are less than
$10^{-4}$. 

We thus see that in this model not only are all three positive 
indications of neutrino oscillations are explained but the mixing between 
the heavier sterile neutrinos $\nu^\prime_{\mu}$ and $\nu^\prime_{\tau}$ 
and the active neotrinos are consistent with all known
oscillation data such as for example the one from the CHOOZ\cite{chooz}. 
What we find very interesting is that with only six parameters describing 
the entire $6\times 6$ neutrino mass matrix (three active and three 
sterile) and every other parameter fixed by the charged 
fermion masses, four neutrino masses and 12 mixing
parameters that link the active to sterile neutrinos which could have 
observable consequences are all completely  consistent with known data.
   
Turning now to the consistency of our model with big bang nucleosynthesis
(BBN), we recall that present observations of Helium and deuterium abundance
can allow for as many as $4.53$ neutrino species\cite{sarkar} if the baryon
to photon ratio is small. There are also a new interesting possibility for
generation of lepton asymmetry in the presence of sterile neutrinos which
will effect the upper limit on the neutrino number\cite{ray}. The 
relevant parameter that determines if extra neutrinos contributes via 
oscillation of the known ones is $\Delta m^2 sin^22\theta \equiv 
\delta_{BBN}$. It has been argued that in the absence of neutrino 
asymmetries, $\delta_{BBN}\leq 10^{-7}~eV^2$.
For large masses of the two mirror neutrinos, this bound is not satisfied.
We therefore invoke the new mechanism proposed in Ref.\cite{ray} 
to evade these bounds thus restoring consistency with the BBN constraints.

The second feature of mirror universe models is the existence of the
mirror photon, which could have experimental manifestations. One arena 
would be the BBN; but as was discussed in Ref.\cite{bere}, this problem
can be ameliorated by the assumption of asymmetric inflation between the 
two sectors\cite{bere2}. A second constraint comes from the fact that 
$\gamma-\gamma^\prime$ kinetic mixing\cite{carlson} is highly constrained by 
the BBN considerations. In our model, since it arises from the mixed 
Higgs field $\chi$ and thus depends on the splittings among the various 
sub-multiplets in the mixed field $\chi$, at the phenomenological level, 
this will serve to
fix the intra-multiplet splitting. It was noted by Carlson and 
Glashow\cite{carlson} that positronium-mirror-positronium oscillation
could also constrain the $\gamma-\gamma^\prime$ mixing. But in our model, we 
assume the mirror weak scale to be about ten times large so that 
${e^+}^\prime{e^-}^\prime$ bound state will have mass of about 10 MeV, 
preventing the possibility of this oscillation.  

In conclusion, we have presented a supersymmetric grand unified model
for the sterile neutrino which can explain the solar, atmospheric and
LSND data. The new features of the model are (a) the use of the type II 
seesaw mechanism to explain the smallness of the neutrino masses while 
at the same time accomodating maximal $\nu_{\mu}-\nu_{\tau}$ oscillation
with degenerate  neutrinos and (b) the prediction of the neutrino mixings
among six light neutrinos (three active and three mirror) with very few
parameters, which are consistent with all known constraints.

This work is supported by the National Science Foundation under
grant no. PHY-9421385. We wish to thank Markus Luty for some discussions.

\begin{table}[htb]
\begin{center}
\[
\begin{array}{|c||c||c||c||c||c|}
\hline
a_1&a_2&a_3&a_4&a_5&a_6 \\
0.140&0.230&0&50.54&-55.30&0 \\
\hline
\hline
c_1&c_2&c_3&c_4&c_5&c_6 \\
0.0016&0.0023&-0.00012&0.579&-0.55&0.0288 \\
\hline
b_1&b_2&b_3&b_4&b_5&b_6 \\
-0.230 & 0 & -0.508 & 0 & -0.139 & 5.64\\
\hline
\hline
d_1&d_2&d_3&d_4&d_5 & d_6\\
0.00076&0.0043&0.0016&0.0096& 0.00037 & -0.0152 \\
\hline
\end{array}
\] 
\end{center}
\caption{The fitted values of $a_i$, $b_i$, $c_i$ and $d_i$}
\label{table}
\end{table}

\end{document}